\documentclass[a4paper,12pt]{article}
\usepackage[dvips]{graphicx}
\usepackage{amsmath,amssymb}
\usepackage{amssymb}
\usepackage{enumerate}
\usepackage{graphicx}
\usepackage{dcolumn}
\usepackage{bm}
\usepackage{bbm}

\usepackage[usenames]{color}

\newcommand {\beq}{\begin{eqnarray}}
\newcommand {\eeq}{\end{eqnarray}}
\newcommand {\non}{\nonumber\\}

\newcommand {\1}[1]{\frac{1}{#1}}

\newcommand {\tr}{{\rm tr}\,}

\newcommand{\vs}[1]{\vspace{#1 mm}}
\newcommand{\hs}[1]{\hspace{#1 mm}}

\newcommand{\p}{\partial}

\newcommand{\bpm}{\begin{pmatrix}}
\newcommand{\epm}{\end{pmatrix}}
\newcommand{\D}{\mathcal D}

\newcommand{\ba}{\left( \begin{array}}
\newcommand{\ea}{\end{array} \right)}

\begin{document}

\begin{titlepage}

\begin{flushright}
{\tt arXiv:0802.1020} \\
IFUP-TH/2008-02\\ 
TIT/HEP-579\\
DAMTP-2008-8
\end{flushright}

\begin{center}

{\LARGE
Constructing Non-Abelian Vortices\\[3mm] with Arbitrary Gauge Groups
}


\vspace{1cm}
Minoru~Eto$^{1,2}$,
Toshiaki~Fujimori$^3$,
Sven~Bjarke~Gudnason$^{1,2}$,
Kenichi~Konishi$^{1,2}$,\\
Muneto~Nitta$^4$,
Keisuke~Ohashi$^5$
and Walter~Vinci$^{1,2}$


\bigskip\bigskip\bigskip
{\it
$^1$  
~Department of Physics, University of Pisa,
Largo Pontecorvo, 3, Ed. C, 56127 Pisa, Italy
\\
$^2$  INFN, Sezione di Pisa,
Largo Pontecorvo, 3, Ed. C, 56127 Pisa, Italy \\
$^3$ Department of Physics, Tokyo Institute of
Technology, Tokyo 152-8551, Japan\\
$^4$ Department of Physics, Keio University, Hiyoshi,
Yokohama, Kanagawa 223-8521, Japan\\
$^5$ Department of Applied Mathematics and Theoretical Physics,
University of Cambridge, CB3 0WA, UK\\
}

\end{center}

\date{\today}

\begin{abstract}
We construct the  general vortex solution in the fully-Higgsed, 
color-flavor locked vacuum of a non-Abelian gauge theory, where the gauge group is taken to
be the product  of
an  arbitrary simple group and
 $U(1)$,  with a Fayet-Iliopoulos term. 
The vortex moduli space is determined. 
\end{abstract}
%
%
\begin{flushright}
\end{flushright}

\end{titlepage}

\newpage

%
%
\section{Introduction}

Vortices play important roles in various areas of physics from  condensed matter physics such as superconductors,  superfluids, quantum Hall effects  to particle physics \cite{Achucarro:1999it} and cosmology  \cite{Vilenkin}.
Recently there has been a significant progress in the understanding of {\it non-Abelian vortices}  in the color-flavor locked vacuum of $SU(N) \times U(1)$ gauge theories  \cite{Hanany:2003hp,Auzzi:2003fs}.
 Unlike Abelian vortices \cite{Abrikosov:1956sx},
they carry orientational moduli  in the internal space, in
addition to the usual position moduli. The most general 
Bogomol'nyi-Prasad-Sommerfield (BPS) vortex solutions and their moduli space have been
found  \cite{Eto:2005yh,Eto:2006pg} and the dynamics of two colliding vortices studied
\cite{Eto:2006db}.  Though these and many other interesting features  have been extensively explored  \cite{Eto:2005yh}-\cite{Eto:2007aw},
most studies so far have  been restricted to the gauge group $SU(N) \times
U(1)$, with  a few but notable  exceptions \cite{Ferretti:2007rp},\cite{DKO}.

We present here a simple framework for writing the most general
non-Abelian BPS vortex solutions in theories with an  arbitrary gauge group of the type  $G=G' \times U(1)$.  For concreteness we take  $G'$ to be a simple Lie group, but the method can be easily generalized to non-simple groups. The cases of classical groups $G'=SU,SO,USp$ will be worked out in detail. Various new physical results seem to follow from our study, even though here we limit ourselves to the main results only.  Fuller account will be given elsewhere\cite{EFGKNOV}.  
\section{Model and BPS Vortex Equations}

We focus our attention on the classical Lie groups $G'=SU(N), SO(N)$ and $USp(2M)$, leaving the exceptional groups to a
short  discussion at the end.
For  $G'= SO(N), USp(2M)$ their group elements are embedded into $SU(N)$ ($N=2M$ for $USp$) by constraints of the form,   $U^T
J U = J$, where $J$ is the rank-2 invariant tensor
\beq
 J = \ba{cc}
            \mathbf{0}_M & \mathbf{1}_{M} \\
 \epsilon \mathbf{1}_{M} & \mathbf{0}_M \ea
 \ ,\quad
 \ba{ccc}
  \mathbf 0_{M} & \mathbf 1_{M} & 0\\
  \mathbf 1_{M} & \mathbf 0_{M} & 0 \\
              0 & 0             & 1\ea \  ,
\eeq
where $\epsilon = +1$ for $SO(2M)$, while $\epsilon = - 1$ for $USp(2M)$; the second matrix is for $SO(2M+1)$.  Apart from the gauge bosons  $W_{\mu} = W_{\mu}^0 t^0 + W_{\mu}^a t^a$  the matter
content of the model consists of  $N$
flavors of  Higgs scalar fields in the fundamental representation, 
 with a common $U(1)$ charge, written as a color-flavor mixed
$N \times N$ matrix $H$.  $t^0$ and $t^a$ denote the generators of $U(1)$ and $G'$
normalized as
\beq
 t^0 = \frac{\mathbf 1_{N}}{\sqrt{2N}} \ , \hs{10}
 \tr ( t^a t^b ) = \frac{\delta^{ab}}{2} \ .
\eeq
The Lagrangian is given by
\beq
\mathcal L &=&- \frac{1}{4e^2} F_{\mu\nu}^0 F^{0\mu\nu}
   - \frac{1}{4g^2} F_{\mu\nu}^a F^{a\mu\nu}
   + \left(\D_\mu H_A\right)^\dag \D^\mu H_A \non
&& - \frac{e^2}{2} \left|H_A^\dagger t^0 H_A - \frac{v^2}{\sqrt{2 N}}\right|^2
   - \frac{g^2}{2} |H_A^\dagger t^a H_A|^2 \ ,
\label{bosonic} \eeq
where
${\cal D}_{\mu} H = (\partial_{\mu} + i W_{\mu})H$,
$e$ and $g$ are the gauge coupling constants
for $U(1)$ and $G'$, respectively, and $A$ is the flavor index.
The flavor symmetry of the model is $SU(N)_{\rm F}$. Though our discussion concerns mainly the bosonic system, (\ref{bosonic}), the model is really to be considered as the (truncated) bosonic sector of the corresponding ${\cal N}=2$ supersymmetric gauge theory, which explains the particular form of the potential,  ensuring at the same time  its stability against radiative corrections. 
The  Bogomol'nyi completion for static, $x^{3}$ independent configurations
\begin{align}
T &= \int d^2 x \bigg[ 
  \frac{1}{2e^2} \left|F_{12}^0
   - e^2 \left(H_A^\dagger t^0 H_A - \frac{v^2}{\sqrt{2 N}}\right) \right|^2  \non
   &\phantom{= \int d^2 x \bigg[}
   + 4\left|\D_{\bar z} H\right|^2
   + \frac{1}{2g^2}  \left|F_{12}^a
   - g^2 H_A^\dagger t^a H_A \right|^2
   - \frac{v^2}{\sqrt{2 N}} F_{12}^0 \bigg] \notag \\
&\geq ~ - \frac{v^2}{\sqrt{2 N}} \int d^2 x \, F^0_{12} \ ,
\end{align}
yields the  BPS vortex equations
\begin{align}
 \D_{\bar z} H &= 0 \ , \label{eq:BPSeq1}\\
 F_{12}^0 - \frac{e^2}{\sqrt{2 N}} \left(\tr(HH^\dagger) - v^2\right)
 & = 0 \ 
  \label{eq:BPSeq2},\\
 F_{12}^a t^a - \frac{g^2}{4}
 \left(HH^\dagger- J^\dagger (HH^\dagger)^T J\right) &= 0 \ ,
  \label{eq:BPSeq3}
\end{align}
for the groups $G'=SO(N), USp(2M)$, where 
a complex coordinate $z \equiv x^1 + i x^2$ has been introduced. Equation (\ref{eq:BPSeq3})
reads for $G'=SU(N)$ instead: 
\beq F_{12}^at^a - \frac{g^2}{2}\, \left[\, HH^\dag -
\frac{\mathbf{1}_N}{N} \tr(HH^\dag)\, \right] = 0 \ . \eeq

\section{Solving the BPS Vortex Equations}
Let us choose the fully Higgsed, color-flavor locked vacuum: $\left< H\right>=\frac{v}{\sqrt{N}}{\bf 1}_N$. 
The $G' \times U(1) \times SU(N)_{\rm F}$ invariance of the theory is broken  to the global color-flavor
diagonal  $G'_{\rm C+F}$. Introduce an $N$ by $N$ matrix $S(z,\bar z)$ taking a value in the
complexification $G^{\mathbb{C}}$ of $G$,
\beq
 S(z,\bar z) = S_e(z,\bar z) S'(z,\bar z) \ , 
\eeq
with $S_e \in U(1)^{\mathbb C} \simeq {\mathbb C}^*$ and $S' \in G^{\prime \mathbb{C}}$.   The  gauge fields can be taken to be equal to the Maurer-Cartan form: 
\beq
   (W^0_1 + i W^0_2) \, t^0
 &=& -  2 i \, S_e^{-1} \bar \partial S_e \ ,\\
(W_1^a + i W_2^a)\,  t^a
 &=& -  2 i \, S^{\prime -1} \bar \partial S^\prime \ .
\eeq
The first of the BPS  equations (\ref{eq:BPSeq1}) can then be solved by 
\beq
 H = S^{-1} H_0 (z) = S_e^{-1} {S'}^{-1}H_0 (z) \ ,
\eeq
where $H_0(z)$ is a matrix whose elements  are holomorphic in 
$z$. $H_0(z)$ will be called moduli matrix
\cite{Isozumi:2004vg}, as all moduli parameters are encoded in it (see below). $H_0(z)$ is   defined up to equivalence relations of the form
\beq
  (H_0, S) \sim V_{e}\,  V^{\prime}(z) (H_0, S) \ , \quad  
   V^{\prime} (z)^{T} J V^{\prime}(z) = J \ .   \eeq
By introducing $N \times N$ matrices $\Omega_0 \equiv H_0 H_0^\dagger$ and
\beq
 \Omega_e \equiv S_e S_e^\dagger \equiv e^{\psi \mathbf 1_{2N}} \in
 U(1)^{\mathbb C} \ , \quad
 \Omega' \equiv S' S^{\prime \dagger} \in {G'}^{\mathbb C} \ ,
\eeq
the BPS equations (\ref{eq:BPSeq2}),
(\ref{eq:BPSeq3}) can be  cast into  the form: 
\begin{align}
  \bar \p \p \psi &= - \frac{e^2}{4N} 
\big(\tr( \Omega_0 {\Omega'}^{-1}) e^{-\psi} - v^2 \big) \ , \\
 \bar \p (\Omega' \p {\Omega'}^{-1})
&= \frac{g^2}{8} \big( \Omega_0 {\Omega'}^{-1} - J^\dagger (\Omega_0
{\Omega'}^{-1})^T J \big) e^{-\psi}\ ,   \nonumber   
\end{align} which we denote the master equations.
The boundary conditions are $\tr(
\Omega_0 \Omega'{}^{-1}) e^{-\psi} = v^2$
and $\Omega_0 \Omega'{}^{-1} = J^\dagger (\Omega_0 {\Omega'}^{-1})^T J$.  We assume the existence and uniqueness for the
solutions to these equations.  There are at least two justifications for this. 
One is the fact that in the strong
coupling limit ($e,g \to \infty$) these can be algebraically and uniquely solved.  The other relies on the index
theorem: the number of the moduli parameters  encoded in $H_0$ coincide
with that obtained from the index theorem \cite{EFGKNOV}. 

The tension of the BPS vortices can be written as
\beq
T = - \frac{v^2}{\sqrt{2 N}} \int d^2 x ~ F^0_{12} = 2 v^2  \int d^2 x
~ \bar \p \p \,  \psi \ .
\eeq
The asymptotic behavior
\begin{eqnarray}
 S_e\sim |z|^{\nu} \quad{\rm for~} |z|\rightarrow \infty \  ,
\end{eqnarray}
then determines the tension
\beq
 T = 2 \pi v^2 \nu \ ,
\eeq
 a rational number $\nu (> 0)$  being the $U(1)$ winding number.  $\nu$ will be found to be 
quantized in half-integers  ($\nu=k/2$) for the groups $G'=SO(2M), USp(2M)$ with $k \in {\mathbb Z}_+$;     $\nu =k$ (integers) 
for $G'=SO(2M+1)$; finally  $\nu=k/N$ for $G'= SU(N)$, as is well known.  The integer $k$ denotes the vortex number:  $k=1$ corresponds to  the minimal vortex in all cases. 

The key idea of this Letter, which enables us to extend the moduli-matrix formalism to  general gauge groups,   is to consider the holomorphic invariants $I_{G'}^i(H)$ made of $H$, which are
invariant under $G'{}^{\mathbb C}$, with $i$ labeling them. If  the $U(1)$ charge of the $i$-th invariant $I_{G'}^i(H)$
is $n_i$,  the following relation 
\begin{eqnarray}
 I^i_{G'}(H)=I_{G'}^i\left(S^{-1}_e{S'}^{-1}H_0\right)
 = S_e^{-n_i}I^i_{G'}(H_0(z))\ ,
 \label{eq:G'inv}
\end{eqnarray}
holds.   If 
the boundary condition is given by 
\begin{eqnarray}
I^i_{G'}(H)\Big|_{|z|\rightarrow \infty} =
 I^i_{\rm vev} \, e^{i \nu n_i \theta}
 \label{eq:boundary}\ ,
\end{eqnarray}
where $\nu \,  n_i$ is 
the number of zeros of $I^i_{G'}$, 
it follows that 
\begin{eqnarray}
 I^i_{G'}(H_0)=S_e^{n_i}I^i_{G'}(H)\sim I^i_{\rm vev} \, z^{\nu \,
   n_i} \ ,
\quad      |z|\rightarrow \infty \  .
\end{eqnarray}
As $I^i_{G'}(H_0(z))$ are holomorphic,
the above condition implies 
that $I^i_{G'}(H_0(z))$ are {\it polynomials} in $z$.
We find that  $\nu\, n_i$ must be a positive integer for all $i$: 
\begin{eqnarray}
\nu \, n_i \in {\mathbb Z}_{+}
\quad \rightarrow \quad \nu = {k}/{n_0} \ , \quad k\in {\mathbb Z}_{+}
\ ,
\end{eqnarray}
where  (GCD  $=$ the greatest common divisor) 
\begin{eqnarray}
 n_0\equiv {\rm GCD}\{n_i~|~I_{\rm vev}^i \neq 0 \} \ .
\end{eqnarray}
Note that a $U(1)$ gauge transformation $e^{2\pi i /n_0}$ leaves  invariant
$I^i_{G'}(H)$:
\begin{eqnarray}
 I^i_{G'}(H')=e^{2\pi i n_i/n_0}I^i_{G'}(H)=I^i_{G'}(H):
\end{eqnarray}
the phase rotation
$e^{2\pi i /n_0}\in {\mathbb Z}_{n_0}$
changes no physics, and  the true gauge group is thus
\begin{eqnarray}
 G={U(1)\times G'}/{{\mathbb Z}_{n_0}}\;.
\end{eqnarray}
where  ${\mathbb Z}_{n_0}$ is the center of $G'$. A simple homotopy argument tells us  that $1/n_0$
is the $U(1)$ winding for the minimal ($k=1$) vortex configuration. 
Finally,  for a given  $k$ the following important relations hold
\begin{eqnarray}
 I_{G'}^i(H_0)=
 I^i_{\rm vev} z^{k n_i/n_0}+{\cal O}(z^{k n_i/n_0-1}) \ ,
 \label{eq:H0const}
\end{eqnarray}
which imply  nontrivial constraints on $H_0(z)$.

The explicit form of the constraints follows from this general discussion.
For   $G'=SU(N)$ (with $N$ flavors),
there exists only one invariant
\begin{eqnarray}
 I_{SU}=\det(H) \ , \label{eq:SUinvariant}
\end{eqnarray}
with charge $N$.
Thus the minimal winding ($1/n_0$)  is equal to  $1/N$ and
the condition for $k$ vortices is given by:
\begin{eqnarray}
 A_{N-1}: \det H_0(z)=z^k+{\cal O}(z^{k-1}), \quad
 \nu = k/N \ .
\label{eq:SUconstraint}
\end{eqnarray}
For  $G'=SO(N),USp(2M)$,
there are  $ N (N \pm 1)/2$ invariants
\begin{eqnarray}
 (I_{SO,USp})^{r}{}_s = (H^{\rm T}J H)^r{}_s,
 \quad 1\le r\le s\le N 
 \label{eq:I-SOUSp} \ ,
\end{eqnarray}
in addition to (\ref{eq:SUinvariant}). The constraints are:
\begin{align}
{C_M,D_M}&: H_0^T(z) J H_0(z) = z^k J + {\cal O}(z^{k-1})\ ,
 \; \ \,\, \nu =k/2 \ , \nonumber\\
{B_M}&: H_0^T(z) J H_0(z) = z^{2k} J + {\cal O}(z^{2k-1})\ ,
 \; \nu = k \ , 
\end{align}
for $G' = SO(2M),USp(2M)$ and
$SO(2M+1)$, respectively.
As anticipated, vortices in the  $SO(2M+1)$ model have integer
$U(1)$ windings \cite{Ferretti:2007rp}.

Explicitly,  the minimal vortices
 in $SU(N)$ and  $SO(2M)$ or $USp(2M)$ theories are given respectively  by the moduli matrices:
\beq
 H_0 =
 \ba{cc}
 z -  a & 0 \\
 {\bf b} & {\bf 1}_{N-1}
 \ea,\
 \ba{cc} z{\bf 1}_M - { \bf A} & {\bf C}_{S/A} \\
         {\bf B}_{A/S}             & {\bf 1}_M
 \ea
 \label{eq:SO,k=1semiloc}.
\eeq
The moduli parameters are all complex. For $SU(N)$, $a$ is just a number;  ${\bf b}$ is a column vector. For $SO(2M)$ or $USp(2M)$, the matrix ${\bf C}_{S/A}$ for instance  is symmetric or antisymmetric, respectively. And vice versa for ${\bf B} $.  Moduli matrices for $SO(2M+1)$ as well as those for $k=2$ vortices in $SU, SO, USp$ theories, 
will be given explicitly  \cite{EFGKNOV}.

The index theorem gives the complex dimension of the
moduli space 
\beq \dim_{\mathbb{C}}\left(\mathcal{M}_{G',k}\right) = \frac{k\, N^2}{n_0} \
. \label{dim_moduli_space_generic} \eeq
 This was obtained  in \cite{Hanany:2003hp} for $SU(N)$; 
a proof in other cases will be reported elsewhere
\cite{EFGKNOV}. In all cases studied  we have checked that the dimension of the moduli space inferred from our moduli matrices agrees with the one given in  Eq.~(\ref{dim_moduli_space_generic}). 

Except for the $SU(N)$ case,  our model has a non-trivial Higgs branch (flat directions).
 The color-flavor locked vacuum  $\langle H\rangle \propto {\bf 1}_N$
is just one of the possible (albeit the most symmetric) choices for the vacuum;  our discussion can readily be
generalized to a generic vacuum on the Higgs branch. This fact, however,  implies that our non-Abelian vortices have ``semilocal'' moduli  (see Achucarro et. al. \cite{Achucarro:1999it}), even for $N_{f}=N$.  In contrast to the Abelian or $SU(N)$ cases, moreover, they exhibit new, interesting phenomena such as ``fractional'' vortices \cite{EFGKNOV}.

\section{Local (ANO-like) Vortices}

For various considerations,   we are interested in knowing  
which of the moduli parameters describe the so-called local vortices,  the ANO-type vortices with  
 exponential tails. 
To identify these,  let us first consider generic  points in the moduli space. In the strong coupling limit  our
theory reduces to a nonlinear sigma model,  with the (classical) vacuum moduli ${\cal M}_{\rm vac}$ as its target
space. In such a limit, semilocal vortices with non-zero size moduli reduce to the so-called sigma model lumps. The local
vortices on the other hand shrink to singular configurations. It is well-known that lumps are characterized by $\pi_2 ({\cal M}_{\rm
vac})$ with a wrapping around a 2-cycle inside ${\cal M}_{\rm vac}$.  Even at a finite gauge coupling, asymptotic
configurations of semilocal vortices can be well approximated by lumps.

Now  the moduli space of vacua ${\cal M}_{\rm vac}$ in supersymmetric models is parametrized by holomorphic invariants
$I_G^I (H)$ ($I=1,2,\ldots$) of the complexified gauge group $G^{\mathbb C}$ \cite{Luty:1995sd}.  In our case, $G=G'
\times U(1)$, with the common $U(1)$ charge of the scalar fields $H$,  all the $G^{\mathbb C}$ invariants $I_G^I (H)$
can be written using the ${G'}^{\mathbb C}$ invariants $I_{G'}^i(H)$. For instance from  $I_{G'}^i$ and
$I_{G'}^j$ with $n_i = n_j$, one can construct
\beq
 I_G^{(i,j)} (H) \equiv \frac{I_{G'}^i(H)}{I_{G'}^j(H)} = 
\frac{I^i_{G'}(H_0(z))}{I^j_{G'}(H_0(z))}
 \label{eq:G'inv/G'inv}\ ,
\eeq
where use was made of (\ref{eq:G'inv}). 
The last line defines, so called, (generalized) rational maps.  
This observation allows us to define local vortices. While asymptotic region of {\it semilocal}
vortices are mapped to some domain of ${\cal M}_{\rm vac}$,
those around the {\it local} vortices are mapped into a single point.  Therefore,  all
the $G^{\mathbb C}$ invariants $I_G^I (H)$ must be constant for the latter. All the $I_{G'}(H)$'s  have zeros at the vortex positions and
winding around them as seen in (\ref{eq:boundary}). These facts,  together with (\ref{eq:G'inv/G'inv}),  imply that all
$I^i_{G'}(H_0(z))$'s with the same $n_i$ must have common zeros:
\beq
 I_{G'}^i(H_{0, {\rm local}})= \left[ \prod_{\ell=1}^{k}   (z-z_{0 \ell})   \right]^{n_{i}/n_{0}} \,   I^i_{\rm vev}\;.   \label{eq:H0constloc} \eeq
For  $G'=SO(2M), USp(2M)$
with $I_{SO,USp}$ of (\ref{eq:I-SOUSp})
we find that the condition for vortices to be of local type is
\beq
 H_{0, {\rm local}}^T(z) J H_{0, {\rm local}}(z) = \prod_{\ell =1}^k
 (z-z_{0 \ell})  \, J \ .
 \label{eq:local-cond}
\eeq

Let us now discuss  a few concrete examples. The general solution for the  minimal  vortex
(\ref{eq:SO,k=1semiloc}) for  $G'=\{SU(N),SO(2M), USp(2M)\}$ is reduced
to a local vortex if we restrict it to be of the
 form:
\beq
 H_{0, {\rm local}} =
 \ba{cc}
 z - a & 0 \\
 {\bf b} & {\bf 1}_{N-1}
 \ea,\
 \ba{cc} (z - a){\bf 1}_M  & 0  \\
         {\bf B}_{A/S}             & {\bf 1}_M
 \ea
 \label{eq:SO,k=1}.
\eeq
The vortex position is given by  $a$.  ${\bf b}$ for $SU(N)$ and ${\bf B}_{A/S}$ for
$SO(2M)$ or $USp(2M)$ encode the Nambu-Goldstone modes associated with the breaking of the color-flavor symmetry by the vortex $G'_{\rm C+F} \to H_{G'}$. The moduli spaces are
direct products of a complex number and the Hermitian
symmetric spaces
\beq
 {\cal M}_{G',k=1}^{\rm local}
 \simeq {\mathbb C} \times G'_{\rm C+F}/H_{G'} \ ,
\eeq
$H_{SU(N)}=SU(N-1)\times U(1)$ while $H_{SO(2M),USp(2M)} = U(M)$. The results for $SU(N)$ and $SO(2M)$ are well-known
 \cite{Hanany:2003hp,Auzzi:2003fs,Ferretti:2007rp}. The matrices (\ref{eq:SO,k=1}) describe just one
patch of the moduli space. In order to define the manifold
globally we need a sufficient number of patches. 
 The number of patches is $N$ for $G'=SU(N)$ and $2^M$ for $G'=SO(2M),USp(2M)$. The transition
functions correspond to the $V$-equivalence relations~\cite{Eto:2005yh,Eto:2006pg}. 
 In the case of $G'=SO(2M)$, 
the patches are given by permutation of the $i$-th and the ($M+i$)-th columns in (\ref{eq:SO,k=1}). 
We find that no regular transition functions connect the odd and even
 permutations (patches), hence the moduli space consists of 
 two disconnected copies of $SO(2M)/U(M)$ \cite{Ferretti:2007rp}. The complex
 dimensions of the moduli spaces are $\dim_{\mathbb
C}{\cal M}_{SO(2M),k=1}^{\rm local}=\1{2} M(M-1) + 1$ and $\dim_{\mathbb
 C}{\cal M}_{USp(2M),k=1}^{\rm local}= \1{2} M(M+1) + 1$.

\section{Exceptional groups}
 
${\boldsymbol E_6}$: There is a rank-3 symmetric tensor: $\Gamma_{ijk}$. The conditions on the moduli matrix is
\begin{eqnarray}
&&
  \Gamma_{i_1i_2i_3}(H_0)^{i_1}{}_{j_1}(H_0)^{i_2}{}_{j_2}
 (H_0)^{i_3}{}_{j_3}
\sim \Gamma_{j_1j_2j_3}z^{k}\ ,
\end{eqnarray}
and the $U(1)$ winding number is quantized as $\nu =k/3$.

${\boldsymbol E_7}$: There are 2 invariant tensors: $d_{ijkl}$ and $f_{ij}$ respectively of rank 4 and 2. The
moduli matrix is constrained as:
\begin{align}
d_{i_1i_2i_3i_4}
 (H_0)^{i_1}{}_{j_1}(H_0)^{i_2}{}_{j_2}(H_0)^{i_3}{}_{j_3}
 (H_0)^{i_4}{}_{j_4}
&\sim
 d_{j_1j_2j_3j_4}z^{2k} \ ,
\non
 f_{i_1i_2} (H_0)^{i_1}{}_{j_1} (H_0)^{i_2}{}_{j_2}
&\sim  f_{j_1j_2} z^{k},
\end{align}
and the vortices are  quantized in half integers: $\nu = k/2$.

${\boldsymbol G_2, \, \boldsymbol F_4, \, \boldsymbol E_8}$ : See Table \ref{table:excepgroups} for the list of the invariant tensors and the
winding numbers.

\section{Conclusion}
We have thus given all the necessary tools to construct vortex solutions in the color-flavor locked vacuum of a
non-Abelian gauge theory with gauge group $G=G'\times U(1)$ where $G'$ is an {\it arbitrary} simple group,
coupled to   Higgs fields in the fundamental representation.
Our method can be extended to other BPS solitons  such as domain walls, monopoles and instantons, and hopefully opens powerful new windows for their investigation.

\smallskip

\begin{table}[ht]
\begin{center}
\begin{tabular}{|c|c|c|c|c|c|c|c|c|}
\hline
$G'$ & $A_{N-1}$& $B_M$ & $C_M, D_M$ &
   $E_6$& $E_7$& $E_8$& $F_4$& $G_2$
 \\ \hline
$R$&
 $N$& $2M+1$& $2M$& $27$& $56$& $248$& $26$& $7$
  \\ \hline
\small{\rm rank\;inv}&
  $-$ & $2$ & $2$ & $3$ & $2,4$ & $2,3,8$ & $2,3$ & $2,3$
 \\ \hline
$n_0$&
  $N$&$1$&$2$&$3$&$2$&$1$&$1$&$1$ \\
\hline
\end{tabular}
\caption{  The dimension of the fundamental
  representation ($R$), the rank of the other invariants
  \cite{Cvitanovic:1976am} 
and  the
  minimal tension ${\cal\nu} = 1/n_0$ i.e.~the center ${\mathbb Z}_{n_0}$ of
  $G'$. 
  The determinant of the $R \times R$ matrix gives one invariant
  with charge, dim $R$.} \label{table:excepgroups} 
\vs{-5}
\end{center}
\end{table}

\section*{Acknowledgment}
\vs{-2} 
 The authors thank Luca Ferretti, Giacomo Marmorini and David Tong for 
useful discussions. 
T.F, M.N. and K.O. would like to thank theoretical 
high energy physics group at Pisa and INFN for their hospitality. 
The work of T.F.~is supported by the Research Fellowships 
of the Japan Society for the Promotion of Science for 
Young Scientists.
The work of M.E.~and K.O.~is also supported by the Research 
Fellowships of the Japan Society for 
the Promotion of Science for Research Abroad. \vs{-3}


%
%


\end{document}